\documentclass[%
floats,aps,showpacs,twocolumn,superscriptaddress,prl,reprint,floatfix,
]{revtex4-1}
\pdfoutput=1

\usepackage{graphicx}

\usepackage{amsmath}

\newcommand{\picending}{pdf}

\newcommand{\ui}{\text{i}}

\newcommand{\flux}{\Phi}
\newcommand{\heff}{h}
\newcommand{\fluxheff}{\flux/\heff}

\newcommand{\mappBachtHK}{\ensuremath{F_\mathrm{kick}}}
\newcommand{\UpBacht}{\ensuremath{U_\mathrm{kick}}}

\newcommand{\mappBdrilled}{\ensuremath{F}}
\newcommand{\UpBdrilled}{\ensuremath{U}}

\newcommand{\maprot}{\ensuremath{F_\mathrm{rot}}}
\newcommand{\Urotation}{\ensuremath{U_\mathrm{rot}}}
\newcommand{\Pebk}{\ensuremath{P_\text{ho}}}
\newcommand{\Uebk}{\ensuremath{U_\text{ho}}}

\newcommand{\qprime}{\ensuremath{q^\prime}}
\newcommand{\pprime}{\ensuremath{p^\prime}}

\newcommand{\pfix}{\ensuremath{{p}_{\text{fix}}}}

\newcommand{\Ach}{A_{\text{ch}}}
\newcommand{\Achone}{A_1} 
\newcommand{\Achtwo}{A_2} 
\newcommand{\Ai}{A_i}

\newcommand{\Nch}{N_{\text{ch}}}

\newcommand{\mulower}{\ensuremath{\mu_\Amulower}}
\newcommand{\muupper}{\ensuremath{\mu_\Amuupper}}
\newcommand{\mui}{\ensuremath{\mu_\Amui}}

\newcommand{\muclassi}{\ensuremath{\mu^{\text{cl}}_\Amui}}
\newcommand{\muclasslower}{\ensuremath{\mu^{\text{cl}}_\Amulower}}
\newcommand{\muclassupper}{\ensuremath{\mu^{\text{cl}}_\Amuupper}}

\newcommand{\atw}{\ensuremath{w_{\infty}}}

\newcommand{\Amuupper}{{B_1}}
\newcommand{\Amulower}{{B_2}} 
\newcommand{\Amui}{{B_i}}

\newcommand{\tHeisenberg}{\ensuremath{\tau_\text{H}}}
\newcommand{\tdwell}{\ensuremath{\tau_\text{d}}}

\newcommand{\Uc}{U_{\text{c}}}

\newcommand{\eem}{w_{12}} 
\newcommand{\aeem}{\langle\eem\rangle}

\begin{document}

\title{Universal Quantum Localizing Transition of a Partial Barrier in a Chaotic Sea}

\author{Matthias Michler}
\affiliation{Institut f\"ur Theoretische Physik, Technische Universit\"at
             Dresden, 01062 Dresden, Germany}

\author{Arnd B\"acker}
\affiliation{Institut f\"ur Theoretische Physik, Technische Universit\"at
             Dresden, 01062 Dresden, Germany}
\affiliation{Max-Planck-Institut f\"ur Physik komplexer Systeme, N\"othnitzer
Stra\ss{}e 38, 01187 Dresden, Germany}

\author{Roland Ketzmerick}
\affiliation{Institut f\"ur Theoretische Physik, Technische Universit\"at
             Dresden, 01062 Dresden, Germany}
\affiliation{Max-Planck-Institut f\"ur Physik komplexer Systeme, N\"othnitzer
Stra\ss{}e 38, 01187 Dresden, Germany}

\author{Hans-J\"urgen St\"ockmann}
\affiliation{Fachbereich Physik, Philipps-Universit\"at Marburg, 35032
  Marburg, Germany}

\author{Steven Tomsovic}
\affiliation{Max-Planck-Institut f\"ur Physik komplexer Systeme, N\"othnitzer
Stra\ss{}e 38, 01187 Dresden, Germany}
\affiliation{Department of Physics and Astronomy, Washington State
  Univeristy, Pullman, Washington 99164-2814, USA}

\date{\today}

\begin{abstract}
  Generic 2D Hamiltonian systems possess partial barriers in their
  chaotic phase space that restrict classical transport. Quantum
  mechanically the transport is suppressed if Planck's constant
  $\heff$ is large compared to the classical flux, $\heff \gg \flux$,
  such that wave packets and states are localized. In contrast,
  classical transport is mimicked for $\heff \ll \flux$. Designing a
  quantum map with an isolated partial barrier of controllable flux
  $\flux$ is the key to investigating the transition from this form of
  quantum localization to mimicking classical transport.  It is
  observed that quantum transport follows a universal transition curve
  as a function of the expected scaling parameter $\fluxheff$. We find
  this curve to be symmetric to $\fluxheff=1$, having a width of two
  orders of magnitude in $\fluxheff$, and exhibiting no quantized
  steps. We establish the relevance of local coupling, improving on
  previous random matrix models relying on global coupling. It turns
  out that a phenomenological $2\times 2$-model gives an accurate
  analytical description of the transition curve.
\end{abstract}

\maketitle

\noindent


In the phase space of generic two-degree-of-freedom (2D) Hamiltonian
systems regions of regular and chaotic motion are dynamically
separated by impenetrable barriers. Within a chaotic region so-called
{\it partial barriers} are ubiquitous. They divide it into distinct
sub-regions, connected by the {\it turnstile
  mechanism}, which works like a revolving door between two rooms. The
volume in phase space, which is transported across the partial barrier
in each direction per time is the flux $\flux$. Partial barriers can
originate \cite{Mei92} from a cantorus or the combination of the
stable and unstable manifold of a hyperbolic fixed point. A hierarchy
of these partial barriers gives rise to a power-law decay of
correlations and of Poincar\'e recurrence time
distributions~\cite{ChiShe1983,*HanCarMei1985,*MeiOtt1985,*CriKet2008}.

What is the implication of a partial barrier on the corresponding
quantum system? In 1984 MacKay, Meiss, and
Percival~\cite{KayMeiPer1984b} conjectured that the flux $\flux$, an
area in phase space, has to be compared with the size $\heff$ of a
Planck cell to judge the quantum implications. For $\heff \gg \flux$
quantum transport is suppressed, while for $\heff \ll \flux$ classical
transport is mimicked
~\cite{BroWya1986,GeiRadRub1986,KayMei1988,BohTomUll1993,MaiHel2000}. Thus, the
existence of a partial barrier in the corresponding classical system
can be conceptualized as being responsible for partially localizing
the quantum dynamics.  As is well known, but still remarkable, quantum
mechanics allows for both the suppression or enhancement of transport
through localization~\cite{And58,Casati79,*Fishman82} or tunneling
phenomena, respectively.

Alternatively one can understand the suppression of transport in the
time domain, where one has the Heisenberg time $\tHeisenberg$ and the
dwell time $\tdwell$ with $\tHeisenberg/\tdwell\sim \fluxheff$. For
$\tHeisenberg\ll\tdwell$ a typical classical orbit of length up to
$\tHeisenberg$ does not cross the partial barrier and stays in the
initial region.  To the extent that the basic semiclassical theory is
valid (neglecting tunneling and diffraction, for example), the
properties of the quantum system are determined by such orbits, and
quantum transport must be suppressed.

\begin{figure}[b!]
     \begin{center}
    \includegraphics{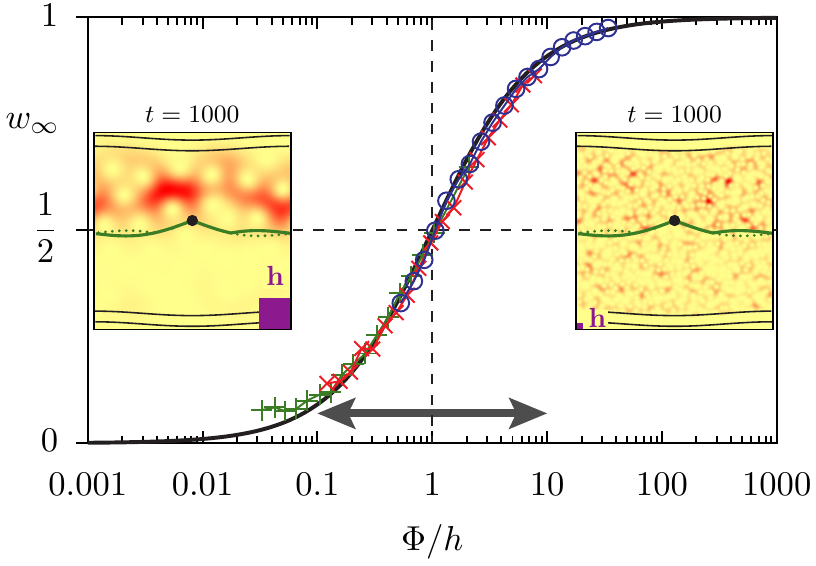}
    \caption{%
      (Color online) Asymptotic transmitted weight $\atw$ vs.
      $\fluxheff$ for $\flux\approx\frac{1}{3000}$ (pluses),
      $\frac{1}{800}$ (crosses), $\frac{1}{200}$ (circles),
      $\heff=\frac{1}{100}, \dots, \frac{1}{6400}$ compared to
      Eq.~\eqref{eq:ATW_heuristic} (solid line). Insets: Husimi
      representation of time-evolved wave packet for $\flux\approx
      \frac{1}{200}$, $\heff=\frac{1}{40}$ and $\heff=\frac{1}{1000}$.
    } \label{fig:show_pB14_atw_mom}
  \end{center}
\end{figure}

The quantum suppression of transport for $\heff \gg \flux$ has
consequences for the time evolution of a localized wave packet
initially associated with a phase space region on one side of the
partial barrier.  It cannot acquire a substantial weight on the other
side of the partial barrier even in the limit of arbitrarily large
times; see Fig.~\ref{fig:show_pB14_atw_mom} left inset. This is
reflected in the eigenstates having much stronger projection in either
one of the two sides; see Fig.~\ref{fig:show_pB14_M_mom} left insets.
In contrast, for $\heff\ll\flux$ wave packets in
the long-time limit as well as eigenstates extend to both regions as
if the partial barrier were not present. This corresponds to the
classical behavior where in the long-time limit chaotic orbits explore
both regions ergodically.

The quantum localizing transition between quantum suppression and
classical transport has been studied theoretically and experimentally
for multiphoton ionization of atoms~\cite{KayMei1988}, cesium atoms in
optical potentials~\cite{VanBalAmmChr1999}, and microcavity
lasers~\cite{ShiLeeKimLeeYanMooLeeAn2008,*YanLeeShiMooLeeKimLeeAn2008,*ShiHarFukHenSasNar2010,*ShiWieCao2011}.
The most extensive study goes back to Bohigas, Tomsovic, and Ullmo
(BTU) on coupled quartic oscillators where seven chaotic regions are
separated by six partial
barriers~\cite{BohTomUll1993,*BohTomUll1990a,*SmiTomBoh1992}. They
model the quantum mechanism of a partial barrier by globally coupled
random matrices with a transition parameter determined by
$\Phi/\hbar$, which were previously used to describe symmetry
breaking~\cite{RosPor1960,*French88,*GuhWei1990}. They found good
agreement for the implications of partial barriers on spectral
statistics and wave packet dynamics.

While the quantum localizing transition is partially understood, the
full quantitative transition even for an isolated partial barrier so
far is not. In particular one is interested in the transition curve,
including its universal scaling, center, width, shape, and whether it
has quantized steps. This is a prerequisite for understanding the
critical essence in quantum mechanics of the turnstile.

In this paper we analyze the quantum localizing transition with the
help of a designed quantum map with an isolated partial barrier of
controllable flux $\flux$. By studying wave packet dynamics and
eigenstate properties, it is found that the transition from quantum
suppression to classical transport is universal with the expected
scaling parameter $\fluxheff$, is symmetric to $\fluxheff=1$, exhibits
no quantized steps, and has a 10\%-90\% width of two orders of
magnitude in $\fluxheff$; for simplicity the results presented are
quoted for a system with only two regions, both of comparable phase
space areas.  It is shown that the critical essence in quantum
mechanics of a turnstile is a local coupling mechanism in contrast to
the previously used global coupling scheme. We give an analytical
description of the universal transition based on a phenomenological
$2\times 2$-model. The findings are confirmed with the more familiar
standard map, which plays an important role in studies of quantum
chaos and localization~\cite{Casati79,*Fishman82}.

\begin{figure}
     \begin{center}
    \includegraphics{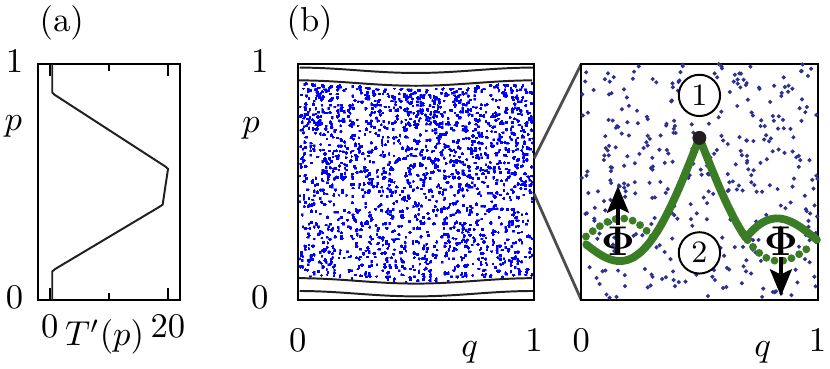}
    \caption{%
      (Color online) (a) $T^\prime(p)$ of the designed map
      $\mappBdrilled$ with $\flux \approx \frac{1}{200}$. (b)
      Corresponding phase space with a large chaotic sea (dots)
      between the regular tori (lines) and zoom with hyperbolic fixed
      point (black circle), partial barrier (solid green line)
      separating regions 1 and 2, its preimage (dotted green line) and
      the turnstile areas of size $\flux$.  }
    \label{fig:show_pB14_atw_mom-2}
  \end{center}
\end{figure}

Consider a family of designed area preserving maps $\mappBdrilled$ of
the two-torus, see Fig.~\ref{fig:show_pB14_atw_mom-2} for an
illustration. The phase space consists of a large chaotic sea between
the regular tori.  There is a hyperbolic fixed point at
$(q,p)=(\frac{1}{2}, \pfix)$, whose stable and unstable manifold can
be used to construct a partial barrier separating the regions 1 and 2
of approximately the same size, $\Achone\approx\Achtwo$.  The region
between the partial barrier and its preimage defines the turnstile
areas of size $\flux$~\cite{Mei92}. The map $\mappBdrilled$ was
designed such that this partial barrier is well isolated with a small
tunable flux. This is achieved by a composition of two maps,
$\mappBdrilled = \maprot \circ \mappBachtHK$. Here $\mappBachtHK$
originates from a kicked Hamiltonian and is given by $(\qprime,
\pprime) = (q + T^\prime(p^*), p^*- V^\prime(\qprime)/2)$ using $p^* =
p - V^\prime(q)/2$ with $V^\prime(q) = \sin(2\pi q)/ (4\pi)$ and
piecewise linear $T^\prime(p)$, see
Fig.~\ref{fig:show_pB14_atw_mom-2}(a). In order to destroy additional
partial barriers not related to the fixed point we use a map
$\maprot$, which rotates points inside a circular region in phase
space specified by center, radius and rotation angle while points
outside this region are unchanged~\cite{Note1}.
The parameters of $\mappBachtHK$ and $\maprot$ for the three
considered cases of $\mappBdrilled$ with fluxes
$\flux\approx\frac{1}{3000}$, $\frac{1}{800}$, and $\frac{1}{200}$ are
given in Refs.~\cite{LongPaper,Mic2011}.
Quantum mechanically the system is described by a unitary operator
$\UpBdrilled=\Urotation\UpBacht$ acting on a Hilbert space of finite
size $N$ with effective Planck's constant $\heff=1/N$. Here $\UpBacht
= \exp\{-\ui V({q}) / (2\hbar)\} \exp\{-\ui T({p}) / \hbar\}
\exp\{-\ui V({q}) / (2\hbar)\}$ and $\Urotation=\Uebk\Pebk +
(\mathbf{1} - \Pebk)$ where $\Pebk$ is a projector built from harmonic
oscillator eigenstates within the rotating region and $\Uebk$ gives the
time evolution corresponding to the classical rotation angle
\cite{Note1}. The time evolution of a wave packet $\psi(t)$ is given
by $\psi(t+1) = \UpBdrilled \psi(t)$.

In order to quantify the quantum transition of a partial barrier we
define the (relative) asymptotic transmitted weight $\atw$ of a wave
packet $\psi_1$ started in region 1 as
\begin{align} \label{eq:def_of_atw} %
  \atw[\psi_1] & := \frac{\langle \mulower[\psi_1(t)] \rangle_t
  }{\muclasslower}.
\end{align}
It is the time-averaged transmitted weight divided by the
corresponding classical weight $\muclasslower$. As initial state we
choose $p_0=0.7$ and as a measure $\mulower$ the probability of
$\psi_1(t)$ within the region $\Amulower$ ($p\in[0.175, 0.325]$). The
classical weight $\muclasslower$ is the relative time a long chaotic
orbit spends in $\Amulower$. Assuming ergodicity in the chaotic sea of
size $\Ach$ it can be determined by $ \muclasslower =
{\Amulower}/{\Ach}$. The definition of $\atw$ implies that it makes a
transition from 0 for $\heff\gg\flux$ to 1 for $\heff\ll\flux$. In the
latter case this happens because any initial state becomes uniformly
distributed at large times as would a classical distribution of
trajectories.

Figure~\ref{fig:show_pB14_atw_mom} shows the resulting $\atw$ vs.
$\fluxheff$ for three different fluxes $\flux$, averaged over 100
values of the Bloch phase and 100 time steps after time $T=2^{20}$.
All data sets fall on top of each other under this scaling, i.e.
$\fluxheff$ is indeed the correct scaling parameter. We expect
that the quantum localizing transition for any partial barrier follows
the same universal curve as a function of $\fluxheff$. This expectation 
assumes that the relative volumes of regions 1 and 2 are of the same 
order and that the mixing time is much shorter than the dwell time.
Figure~\ref{fig:show_pB14_atw_mom} shows that on a logarithmic scale
the transition is symmetric with respect to the point $\fluxheff=1$,
$\atw=\frac{1}{2}$. Thus the transition point is reached when flux
$\flux$ and Planck's constant $\heff$ are equal. The transition is
found to be smooth with no indications for quantized steps at integer
values of $\fluxheff$.  By defining the transition region as the
interval in $\fluxheff$ for which $\atw\in [0.1, 0.9]$, the width is
seen to be two orders of magnitude, indicated by the arrow in
Fig.~\ref{fig:show_pB14_atw_mom}. The overall behavior of the
transition is well described by the symmetric curve of
Eq.~\eqref{eq:ATW_heuristic}, which results from a phenomenological
$2\times 2$ matrix model, see below.  For the smallest $\fluxheff$
(below 0.05)
there is an upward trend of $\atw$ compared to the symmetric curve.
We attribute this deviation to the effect of tunneling across the
entire partial barrier, which occurs in addition to quantum transport
through the turnstile region.  This is left for future investigation.

Complementary to the time evolution of wave packets, consider
properties of eigenstates of the quantum map. States are either
contained in region 1 or in region 2 in the case of $\heff\gg\flux$, see
Fig.~\ref{fig:show_pB14_M_mom} left inset. For $\heff\ll\flux$ the
states extend over the whole chaotic region, see
Fig.~\ref{fig:show_pB14_M_mom} top right inset. In addition there
exist regular eigenstates localized on the invariant tori and scarred
states, e.g.\ on the hyperbolic fixed point.
We define the average eigenstate equipartition measure $\aeem$ by
\begin{align} \label{eq:def_of_average_eem} %
  \aeem & := \frac{1}{\Nch} \sum\limits_{j=0}^{N-1}
  \frac{\muupper[\phi_j]}{\muclassupper}
  \frac{\mulower[\phi_j]}{\muclasslower}
\end{align}
as the sum of the products of the relative weights of eigenstates
$\phi_j$ in each measuring box $\Amuupper$ ($p\in [0.675, 0.825]$) and
$\Amulower$ ($p\in[0.175, 0.325]$) contained in regions 1 and 2, resp. The
relative weight compares the measure $\mui[\phi_j]$ with the
corresponding classical weight $\muclassi$. Eigenstates which are
almost zero in one of the regions give no contribution, while the
contribution is 1 for eigenstates, which are almost equipartitioned.
The prefactor $1/\Nch$ is chosen such that in the semiclassical
limit $\aeem$ approaches 1, as in this limit all $\Nch=\Ach/\heff$
chaotic states contribute 1, while regular states give no contribution.
\begin{figure}[b]
     \begin{center}
    \includegraphics{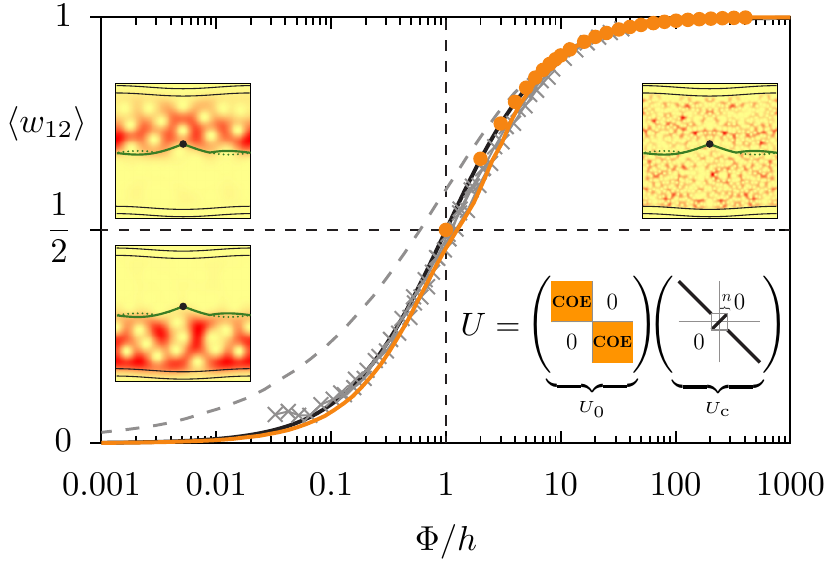}
    \caption{%
      (Color online) Average eigenstate equipartition measure $\aeem$
      for the designed map (gray crosses, same parameters as in
      Fig.~\ref{fig:show_pB14_atw_mom}) compared to
      Eq.~\eqref{eq:ATW_heuristic} (solid line), the BTU model (gray
      dashed line) and the channel coupling model with discrete
      (orange circles) and continuous transmissions (light orange
      line) using 1000 realizations with $N_1=N_2=500$. Inset: Husimi
      representation of typical eigenstates for $\flux\approx
      \frac{1}{200}$, $\heff=\frac{1}{40}$ (left) and
      $\heff=\frac{1}{1000}$ (top right). Illustration of the matrix
      structure of $U_0$ and $\Uc$, where black lines represent unit
      entries (bottom right).
    } \label{fig:show_pB14_M_mom}
  \end{center}
\end{figure}
Figure~\ref{fig:show_pB14_M_mom} shows $\aeem$ for three different
fluxes $\flux$. One observes the same transitional behavior as for
$\atw$, again well described by Eq.~\eqref{eq:ATW_heuristic}. In fact,
one can show that $\aeem=\langle\atw[\psi_1^k]\rangle_k$ if $\atw$ is
averaged over initial states $\psi_1^k$ which form an orthonormal basis
in $\Amuupper$.

In order to phenomenologically describe the transition we propose a
unitary $2\times 2$ matrix model
\begin{equation} \label{eq:U_deterministic_2by2}
  U = 
  \begin{pmatrix} \sqrt{1-v^2} & v \\
                  v & -\sqrt{1-v^2} \\
  \end{pmatrix} .
\end{equation}
Here the deterministic variable $v \in [0, 1]$ describes the turnstile
coupling between two sites representing the chaotic regions $\Achone$
and $\Achtwo$. Following from unitarity the diagonal entries have
magnitude $\sqrt{1-v^2}$. The lower entry has a minus sign such that
for $v=0$ the eigenvalues are not degenerate. The eigenvalues are $\pm
1$ independent of $v$ and the normalized eigenvectors are $\phi_\pm =
1/\sqrt{2} (\pm c_\pm, v/c_\pm)$ with $c_\pm=\sqrt{1 \pm
  \sqrt{1-v^2}}$. According to Eq.~\eqref{eq:def_of_average_eem} the
average eigenstate equipartition measure is $\aeem= v^2$, where the
quantum measures $\mui[\phi_\pm]$ are given by the squared $i$-th
element of the eigenvectors $\phi_\pm$ and the classical expectations
are $\muclassupper=\muclasslower=\frac{1}{2}$.  For the asymptotic
transmitted weight $\atw$, Eq.~\eqref{eq:def_of_atw}, i.e.\ for a wave
packet started on one site and measured on the other site, we find the
same result, $\atw = v^2$.  The parameter $v$ of this model can be
related to the scaling parameter $\fluxheff$ of quantum maps with a
partial barrier: We identify the transmission probability $v^2$ with
the relative classical flux $\flux/\Ai$, i.e.\ $v^2=\flux/\Ai$.
Moreover, we choose the Planck cell associated with each site of the
$2\times 2$-model to be the sub-region of $\Ai$ that is not
transmitted, $\heff=\Ai - \flux$. This choice makes the regions
associated with $\heff$ and $\flux$ disjoint and thereby allows for
arbitrary ratios of $\fluxheff$. This finally gives for the $2\times
2$-matrix model
\begin{align} \label{eq:ATW_heuristic} %
  \atw & = \aeem = \frac{\displaystyle \fluxheff}%
  {\displaystyle 1 + \fluxheff} .
\end{align}
Quite amazingly this phenomenological model gives a very good
description of the transitional behavior of the map data for the
entire range from $\fluxheff \ll 1$ to $\fluxheff \gg 1$ (apart from
the deviation for $\fluxheff\leq0.05$ attributed to tunneling at small $\fluxheff$), see
Figs.~\ref{fig:show_pB14_atw_mom} and \ref{fig:show_pB14_M_mom}.  This
is reminiscent of the success of the Wigner surmise using $2\times 2$
matrices to describe universal spectral statistics.  As no system
specific properties were used in the derivation of
Eq.~\eqref{eq:ATW_heuristic}, except for the scaling parameter
$\flux/\heff$, this gives further support for the
universality of the transition curve. We expect that the universality
also extends to time continuous Hamiltonian systems like billiards.

In order to get an insight into the quantum mechanism of a partial
barrier we now study appropriately adapted random matrix models. In
the BTU model~\cite{BohTomUll1993} two matrices of the Gaussian
orthogonal ensemble, representing two chaotic regions, are globally
coupled with a strength determined by $\fluxheff$.
Figure~\ref{fig:show_pB14_M_mom} shows that for $\fluxheff<10$ this
model overestimates the value of $\aeem$ found for the quantum map.
We attribute this discrepancy to the global coupling of the BTU model.
Instead we propose to model the classical turnstile mechanism by a
local coupling via a channel. In a unitary model $U = U_0 \cdot \Uc$
we decompose the dynamics into a coupling matrix $\Uc$ modeling the
turnstile transport multiplied by an uncoupled matrix $U_0$ modeling
the mixing in each of the regions 1 and 2, see
Fig.~\ref{fig:show_pB14_M_mom} inset.  The coupling matrix $\Uc$ is an
identity matrix, where the central $2n\times 2n$ block has ones on the
anti-diagonal. It couples the two regions via $n$ modes for each
direction of the channel. This models the directed transport of a
turnstile in the classical system. The matrix $U_0$ is block diagonal
consisting of two matrices of the circular orthogonal ensemble
of sizes $N/2 \times N/2$ and the limit of large $N$ is considered.
This model has only one parameter, namely the number of modes
$n=\flux/\heff$.  The resulting $\aeem$ for this unitary channel
coupling model is shown in Fig.~\ref{fig:show_pB14_M_mom} and is in
very good agreement with the numerical data of the map
$\mappBdrilled$. It is stressed that this agreement has been obtained
without any fitting parameter. The model can be extended to
non-integer values of $\fluxheff$ by the continuous transmissions of
modes propagating through a channel~\cite{LongPaper}, see Fig. 3. We
observe that these models with local coupling are better describing
the data for the map $\mappBdrilled$ for $\fluxheff\leq 10$ compared
to the global coupling BTU model and on the same level as the
phenomenological $2\times2$-model. This suggests that the turnstile
mechanism of the classical system is quantum mechanically described by
a local coupling via a channel.

\begin{figure}[b]
  \begin{center}
    \includegraphics{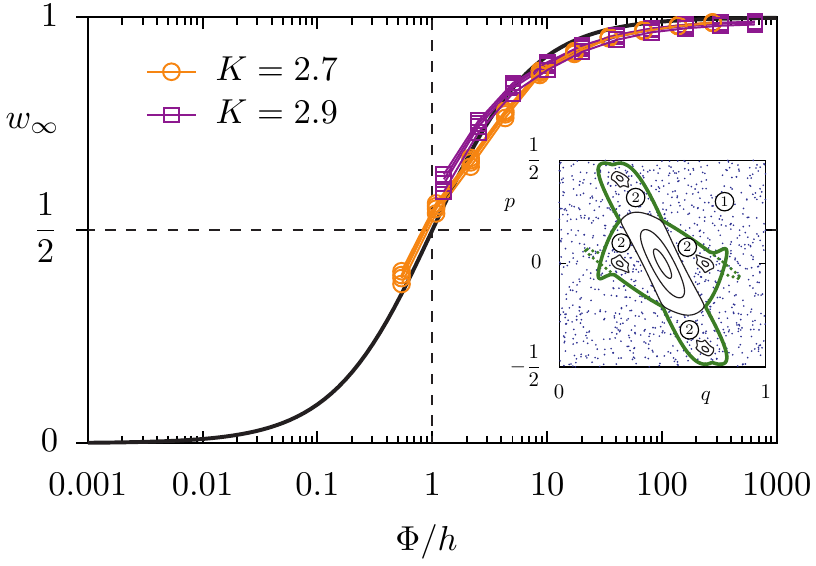}
    \caption{%
      (Color online) Asymptotic transmitted weight $\atw$ using the
      Husimi measure for the standard map with $K=2.7$
      ($\flux\approx\frac{1}{200}$) and $K=2.9$
      ($\flux\approx\frac{1}{100}$) for $\heff=\frac{1}{100}, \dots,
      \frac{1}{51200}$ versus $\fluxheff$ compared to
      Eq.~\eqref{eq:ATW_heuristic} (solid line). The data is averaged
      over $15$ initial conditions placed in region 1, $10$ values of
      the Bloch phase, and $100$ time steps after time $10^6$. Data
      for four different variants~\cite{Mic2011,LongPaper} of the
      partial barrier are shown. Inset: One partial barrier (thick
      solid line) separating regions 1 and 2 for the standard map with
      $K=2.7$ as well as its preimage (dotted lines)
    } \label{fig:show_std_map_atw}
  \end{center}
\end{figure}
We now show results for the generic standard map, $(\qprime, \pprime)
= (q + p, p + K \sin(2\pi \qprime) / (2\pi))$, with kicking strengths
$K=2.7$ and $K=2.9$, where one has a dominant partial barrier
separating two sufficiently large chaotic regions, see
Fig.~\ref{fig:show_std_map_atw} inset and
Ref.~\cite{LongPaper,Mic2011} for details.  We consider the asymptotic
transmitted weight $\atw$ of a wave packet initially located outside
of the partial barrier in region 1. We integrate the Husimi function
of the wave packet $\psi_1(t)$ over the measuring box $\Amulower$,
which we choose as the entire region 2.
Figure~\ref{fig:show_std_map_atw} shows good agreement of $\atw$ over
the accessible range $\fluxheff\geq\tfrac{1}{2}$ with the universal
behavior as observed in Figs.~\ref{fig:show_pB14_atw_mom} and
\ref{fig:show_pB14_M_mom} and well described by
Eq.~\eqref{eq:ATW_heuristic}.

We believe that the understanding of the universal behavior of the
quantum localizing transition of an isolated partial barrier provides
the building block to explain the power--law scaling
\cite{GeiRadRub1986,GreFisPra1984,FisGrePra1987} 
occurring in the presence of hierarchically
organized partial barriers around a cantorus.  Our results also open
the possibility to tackle the tunneling regime which sets in for
$\fluxheff\lesssim 0.1$.  There in addition to the local channel
coupling mechanism of the turnstile one has tunneling across the
entire partial barrier.  Finally, if an extended chaotic system has an
infinite chain of well isolated partial barriers, then the classical
dynamics is diffusive, and the quantum dynamics will lead to
exponential localization no matter how open the partial barriers.
This similarity to Anderson localization \cite{And58} would be very
interesting to explore based on the universal transition curve of
quantum transport through a single partial barrier.

\begin{acknowledgments}
  We are grateful to T.~Guhr, M.~K\"orber, J.~Kuipers, and U.~Kuhl for
  stimulating discussions and acknowledge financial support through
  the DFG Forschergruppe 760 ``Scattering systems with complex
  dynamics.''
  S.~T.\ gratefully acknowledges a Fulbright Fellowship and financial support
  from the US National Science Foundation grant PHY-0855337.
\end{acknowledgments}

\end{document}